\documentclass{iaus}
\usepackage{graphicx}

\title[Galactic evolution of D, {\boldmath $^3$}He and {\boldmath $^4$}He] 
{Galactic evolution of D, $^3$He and $^4$He}

\author[D. Romano]   
{Donatella Romano}

\affiliation{Dept. of Astronomy, Bologna University, \\ Via Ranzani 1, 
I-40127, Bologna, Italy \\ and
\\[\affilskip]INAF-Bologna Observatory, \\ Via Ranzani 1, I-40127, Bologna, 
Italy \\ email: {\tt donatella.romano@oabo.inaf.it}}

\pubyear{2010}
\volume{268}  
\jname{Light elements in the Universe}
\editors{C. Charbonnel, M. Tosi, F. Primas \& C. Chiappini, eds.}
\begin{document}

\maketitle

\begin{abstract}
The uncertainties which still plague our understanding of the evolution of the 
light nuclides D, $^3$He and $^4$He in the Galaxy are described. Measurements 
of the local abundance of deuterium range over a factor of 3. The observed 
dispersion can be reconciled with the predictions on deuterium evolution from 
standard Galactic chemical evolution models, if the true local abundance of 
deuterium proves to be high, but not too high, and lower observed values are 
due to depletion onto dust grains. The nearly constancy of the $^3$He abundance 
with both time and position within the Galaxy implies a negligible production 
of this element in stars, at variance with predictions from standard stellar 
models which, however, do agree with the (few) measurements of $^3$He in 
planetary nebulae. Thermohaline mixing, inhibited by magnetic fields in a small 
fraction of low-mass stars, could in principle explain the complexity of the 
overall scenario. However, complete grids of stellar yields taking this 
mechanism into account are not available for use in chemical evolution models 
yet. Much effort has been devoted to unravel the origin of the extreme 
helium-rich stars which seem to inhabit the most massive Galactic globular 
clusters. Yet, the issue of $^4$He evolution is far from being fully settled 
even in the disc of the Milky Way.

\end{abstract}

\firstsection 
\section{Introduction}
\label{sec:intro}

The discovery of the cosmic microwave background (CMB) by \cite{pw65} in the 
mid sixties set the stage for a quantitative exploitation of the Big Bang 
nucleosynthesis theory. In their pioneer studies, \cite{p66} and \cite{wfh67} 
demonstrated that D, $^3$He and $^4$He could well have been produced in 
solar-system abundances in the `primordial fireball'.

\begin{figure}
  \begin{center}
  \includegraphics[height=10cm]{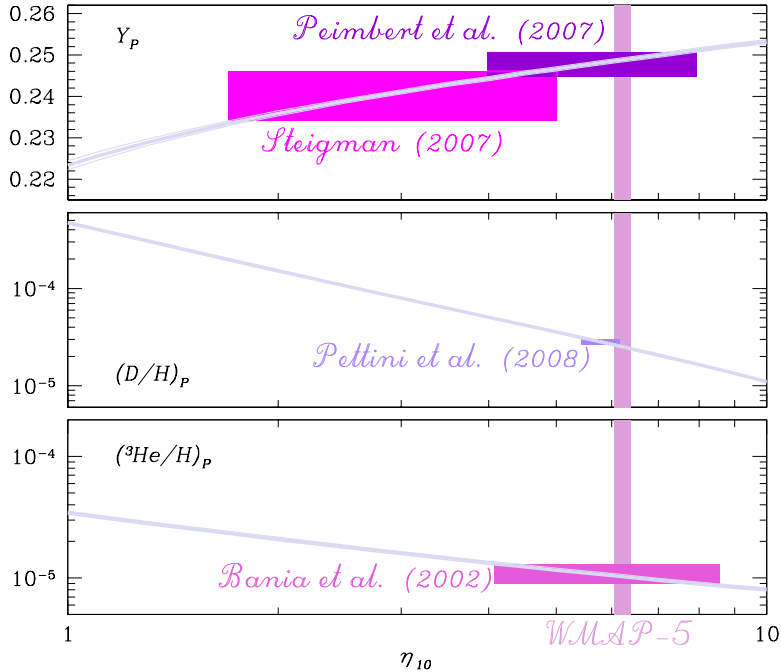} 
  \caption{SBBN-predicted primordial mass fraction of $^4$He ($Y_{\mathrm{P}}$) 
    and abundances of D and $^3$He (relative to hydrogen by number), as 
    functions of the $\eta_{\mathrm{10}}$ parameter, $\eta_{\mathrm{10}} \equiv 
    10^{10} (n_{\mathrm{B}}/n_{\mathrm{\gamma}})$. Theoretical predictions are 
    from \cite{h95}, as updated by G.~Steigman (courtesy of G.~Steigman). The 
    widths of the curves reflect the uncertainties in the nuclear and 
    weak-interaction rates. The vertical band crossing all panels corresponds 
    to the $\eta_{\mathrm{10}}$ value derived from analysis of five-years 
    \emph{WMAP} data on the CMB anisotropy (\cite{d09}). Also shown are the 
    most recent estimates of the primordial abundances of D, $^3$He and $^4$He 
    from observations, along with the allowed ranges of values for 
    $\eta_{\mathrm{10}}$ (boxes; \cite{b02,p07,s07,p08}).}
  \label{fig:prim}
 \end{center}
\end{figure}

In the framework of the standard Big Bang nucleosynthesis (SBBN) theory, the 
baryon-to-photon ratio, $\eta$, is the only parameter regulating the amounts of 
D, $^3$He and $^4$He which emerge from the hot, early Universe (see 
Fig.~\ref{fig:prim} and contribution by G.~Steigman, this volume). In the 
nineties, observations, seeking to constrain the primordial abundances of D, 
$^3$He and $^4$He -- and, hence, the value of $\eta$ -- by probing the most 
metal-poor environments in the Universe, did not come up with consistent 
results. Both high and low values were suggested for the primordial deuterium 
abundance as measured in high-redshift, low-metallicity quasar absorption-line 
systems (QSOALS), (D/H)$_{\mathrm{P}}$~= 2.5~$\times$ 10$^{-4}$ (\cite{c94,s94}) 
or a few times 10$^{-5}$ (\cite{bt98a},\cite{bt98b}). Similarly, both low and 
high values were suggested for the primordial $^4$He abundance, e.g., 
$Y_{\mathrm{P}}$ = 0.234~$\pm$ 0.002 (\cite{o97}) or 0.244~$\pm$ 0.002 
(\cite{it98}).

The difficulty to determine (D/H)$_{\mathrm{P}}$ from observations led to turn 
the problem upside down and try to infer that quantity by using Galactic 
chemical evolution (GCE) models. GCE models put stringent limits on the degree 
of astration suffered by deuterium in the solar vicinity over a Hubble time. 
Assuming that the local pre-solar (\cite{gr72,gg98}) and current (\cite{l98}) D 
abundances are reasonably well known, they could settle tight limits to the 
primordial deuterium abundance, and definitively ruled out a high primordial 
deuterium.

\begin{figure}
  \begin{center}
  \includegraphics[height=10cm]{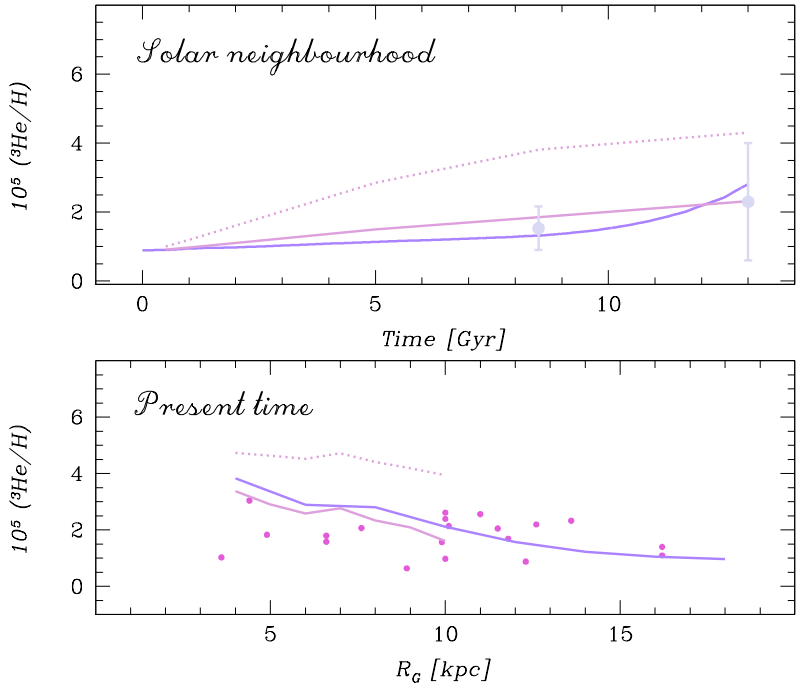} 
  \caption{Evolution of $^3$He/H in the solar neighbourhood (upper panel) and 
    distribution of $^3$He/H across the Galactic disc at the present time 
    (lower panel) for different GCE models, assuming either 
    (D/H)$_{\mathrm{P}}$~= 2.5~$\times$ 10$^{-5}$ (solid lines) or 
    (D/H)$_{\mathrm{P}}$~= 20~$\times$ 10$^{-5}$ (dotted lines). All models 
    assume zero net production of $^3$He from 93\% of low-mass stars (1--2 
    M$_\odot$) in order to fit the observations (filled circles; upper panel: 
    local pre-solar and current values from \cite{gg98}; lower panel: 
    H\,{\small II} region abundances from \cite{b02}). Figure adapted from 
    Romano et al. (2003).}
  \label{fig:3he}
 \end{center}
\end{figure}

Modelling the Galactic evolution of deuterium is a straightforward task. Since 
D is completely destroyed as gas cycles through stars and there are no known 
sources of substantial production other than BBN (\cite{e76,pf03}), its 
evolution is obtained for free from GCE models. Good models for the solar 
neighbourhood -- i.e., models which satisfy the majority of the observational 
constraints available for the solar neighbourhood -- have always predicted 
astration factors $f_{\mathrm{D}} \equiv$ 
(D/H)$_{\mathrm{P}}$/(D/H)$_{\mathrm{LISM}}$ not in excess of 2--3 for deuterium 
(\cite{at74,st92,e94,g95,p96,t98,c02,r03}, 2006). Attempts to accommodate 
larger astration factors (Vangioni-Flam et al. 1994, \cite{s97}) have resulted in models which failed 
to reproduce important observational constraints. Moreover, the more D is burnt 
in the Galaxy, the more $^3$He is produced. Since the abundance of $^3$He is 
observed to stay rather constant with both time and position in the Milky Way 
(\cite{b02}), GCE models which overproduce this isotope must be discarded.

As first recognized by \cite{tc71}, GCE models adopting standard prescriptions 
for the synthesis of $^3$He in stars dramatically overestimate its abundance in 
the Milky Way (see also \cite{r76}). According to standard stellar models, 
$^3$He is most efficiently produced on the main sequence (MS) of 1--2~M$_\odot$ 
stars through the action of the p-p chains. In order not to overproduce $^3$He 
in the course of Galactic evolution, it has become customary to assume that 
some unknown $^3$He-destruction mechanism is at work in more than 90\% of 
low-mass stars (\cite{d96,g97,c02,r03}). \cite{ho95} and \cite{c95} have 
suggested `extra mixing' during the red giant branch (RGB) phase of low-mass 
stars as a possible solution (see also \cite{cdn98}, Sackmann \& Boothroyd 1999). In 
Fig.~\ref{fig:3he}, we compare the predictions of two successful models for the 
chemical evolution of the Milky Way (the one by \cite{c02} and Model~1 of 
\cite{t88}) to $^3$He data for the solar neighbourhood (upper panel) and the 
Galactic disc (lower panel). Despite different assumptions about the infall 
law, star formation rate and stellar initial mass function (IMF), both models 
need to assume that at least 93\% of low-mass stars burn the $^3$He they have 
produced on the MS in later evolutionary phases in order to fit the 
observations. It is worth noticing that the good agreement between model 
predictions and observations depends also on the adopted value of the 
primordial deuterium abundance: if (D/H)$_{\mathrm{P}}$~= 20~$\times$ 10$^{-5}$, 
rather than 2.5~$\times$ 10$^{-5}$ (dotted versus solid lines in 
Fig.~\ref{fig:3he}), both the local behaviour of $^3$He with time and its 
present distribution across the Galactic disc can not be reproduced by the 
models, independently of how many low-mass stars burn their $^3$He on the RGB.

As far as $^4$He is concerned, there has been a general consensus that the 
relative helium-to-metal enrichment ratio in the solar neighbourhood is 
$\Delta Y/\Delta Z \sim$ 2, both from a theoretical (\cite{cm82,m92,c03}) and 
an observational point of view (e.g., \cite{c07}). Yet, hints for very 
different values of this ratio were reported early on in the literature 
(\cite{d70}, and references therein).

The determination of the parameter $\eta$ from \emph{WMAP} data (see text by 
J.~Dunkley, this volume) has allowed to fix, with unprecedented precision, the 
primordial abundances of the light elements in the framework of the SBBN model 
(see \cite{c08} for recent work). The primordial abundances of D, $^3$He and 
$^4$He determined indirectly from the CMB anisotropies agree very well with 
those inferred from recent, direct observations (see Fig.~\ref{fig:prim} and 
Table~\ref{tab:abun}), although one must be aware that the latter actually 
provide only lower/upper limits to the true primordial abundances. Above all, 
it is clear that the determination of the primordial abundance of deuterium is 
converging towards a low value, beautifully confirming earlier findings from 
GCE models.

In the following sections, the remaining (major) causes of uncertainty, which 
hamper our current understanding of the Galactic chemical evolution of the 
light elements D, $^3$He and $^4$He, are discussed, element by element.
 
\begin{table}
  \begin{center}
    \caption{Abundances of D, $^3$He and $^4$He at different epochs}
    \label{tab:abun}
    {\scriptsize
      \begin{tabular}{c c c c c c}
        \hline 
        {\bf Nuclide} & {\bf Units} & {\bf SBBN+\emph{WMAP}}{\boldmath$^a$} & {\bf Low-\emph{Z} systems} & {\bf Pre-solar matter} & {\bf LISM} \\ 
        & & {\bf (13.7 Gyr ago)} & {\bf (10--13 Gyr ago)} & {\bf (4.5 Gyr ago)} & {\bf (Today)} \\
        \hline
        D & 10$^5$ (D/H) & 2.49 $\pm$ 0.17$^{(1)}$ & 2.8 $\pm$ 0.2$^{(2)}$ & 2.1 $\pm$ 0.5$^{(3)}$ & 2.31 $\pm$ 0.24$^{(4)}$ \\
        & & & & & 0.98 $\pm$ 0.19$^{(5)}$ \\
        & & & & & 2.0 $\pm$ 0.1$^{(6)}$ \\
        $^3$He & 10$^5$ ($^3$He/H) & 1.00 $\pm$ 0.07$^{(1)}$ & 1.1 $\pm$ 0.2$^{(7)}$ & 1.5 $\pm$ 0.2$^{(3)}$ & 2.4 $\pm$ 0.7$^{(8)}$ \\
        $^4$He & $Y$ & 0.2486 $\pm$ 0.0002$^{(1)}$ & 0.2477 $\pm$ 0.0029$^{(9)}$ & 0.2703$^{(11)}$ & \\
        & & & 0.240 $\pm$ 0.006$^{(10)}$ & \\
        \hline
      \end{tabular}
    }
  \end{center}
  \vspace{1mm}
  \scriptsize{
    {\it Notes:}\\
    $^a$ Using $\eta_{10}$~= 6.23 $\pm$ 0.17 from analysis of 5-years 
    \emph{WMAP} data (\cite{d09}).\\
    {\it References:}\\
    (1) Cyburt et al. (2008); (2) Pettini et al. (2008); (3) Geiss \& Gloeckler (1998); (4) Linsky et al. (2006); (5) H\'ebrard et al. (2005); (6) Prodanovi\'c et al. (2009); (7) Bania et al. (2002); (8) Gloeckler \& Geiss (1996); (9) Peimbert et al. (2007); (10) Steigman (2007); (11) Asplund et al. (2009).}
\end{table}

\section{Deuterium}
\label{sec:deu}

The joint determinations of the primordial and pre-solar deuterium abundances 
(Table~\ref{tab:abun}) point to a small depletion of deuterium from the Big 
Bang up to the solar system formation 4.5 Gyr ago. However, the present-day 
abundance of deuterium in the solar vicinity is currently under debate. The 
\emph{FUSE} satellite has measured the deteurium abundance along the lines of 
sight to several stars in the Local Bubble as well as beyond it, up to 1--2~kpc 
away. The dispersion (by a factor of 3) which has been found in the 
measurements (\cite{l06}) makes it hard to interpret the data in the context of 
standard GCE models. It has been suggested (\cite{h05,l06}) that either the 
lowest (see also text by G.~H\'ebrard, this volume) or the highest (see also 
text by J.~Linsky, this volume) observed abundances are indicative of the 
actual value of the deuterium abundance in the local ISM (LISM). However, 
neither of these values can be reproduced by GCE models in agreement with all 
the major observational constraints for the solar neighbourhood (\cite{r06}). 
Very recently, using a Bayesian analysis approach, Prodanovi\'c et al. (2009) 
have provided another estimate of the true LISM deuterium abundance, which 
places it very close to the D abundance at the time of the formation of the Sun 
(see Table~\ref{tab:abun}).

\begin{figure}[b]
  \begin{center}
  \includegraphics[height=10cm]{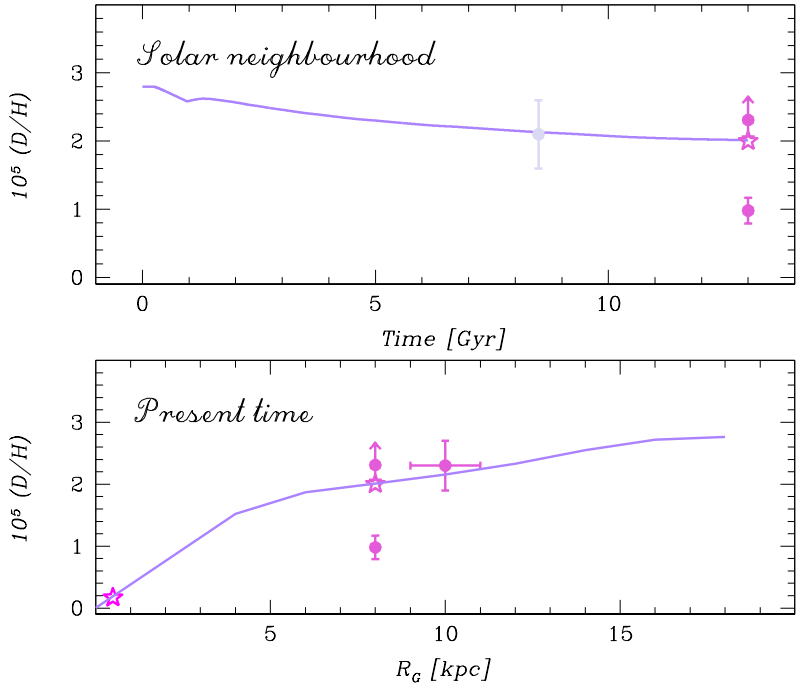} 
  \caption{Evolution of D/H in the solar neighbourhood (upper panel) and 
    distribution of D/H across the Galactic disc at the present time 
    (lower panel) for the GCE model (solid lines in both panels) of Romano et 
    al. (2006) with the lowest D astration factor, assuming 
    (D/H)$_{\mathrm{P}}$~= 2.8~$\times$ 10$^{-5}$. Data are from 
    Table~\ref{tab:abun} for local values, from \cite{r05} for the outer disc 
    (filled circle at $R_{\mathrm{G}}$~= 10~$\pm$ 1 kpc, bottom panel) and from 
    \cite{l00} for the inner Galaxy (star at $R_{\mathrm{G}}$~= 10 pc, bottom 
    panel).}
  \label{fig:deu}
 \end{center}
\end{figure}

In Fig.~\ref{fig:deu} we show the evolution of deuterium in the solar vicinity 
and the deuterium abundance profile across the Milky Way disc. The adopted GCE 
model, from Romano et al. (2006), is the one which allows for the lowest D 
astration factor in the solar vicinity ($f_{\mathrm{D}}$~= 1.39). Here it has 
been recomputed assuming (D/H)$_{\mathrm{P}}$~= 2.8~$\times$ 10$^{-5}$ rather 
than 2.6~$\times$ 10$^{-5}$ as in Romano et al. (2006). Extrapolation of the 
theoretical results towards the Galaxy center has been performed by taking into 
account the detailed results of a model for the Galactic bulge by \cite{m99}. 
Data are from Table~\ref{tab:abun} for local values (at a radius 
$R_{\mathrm{G}}$~= 8 kpc), from \cite{r05} for the outer disc and from 
\cite{l00} for a region at 10 pc distance from the Galactic center.

The agreement of the model predictions with the data is striking, especially if 
the true value for the local abundance of deuterium is the one suggested by 
Prodanovi\'c et al. (2009; see also contribution by T.~Prodanovi\'c, this 
volume). In this context, lower observed local abundances of deuterium would be 
due to D depletion onto dust grains (\cite{l06,srt07}, and references therein). 
The highest observed local D values are marginally consistent with the estimate 
of the true D value by Prodanovi\'c et al. (2009).

\section{Helium-3}
\label{sec:3he}

As far as $^3$He is concerned, we must recognize that the problems raised in 
pioneering works by people such as \cite{tc71}, \cite{r73} and Tinsley (1974) (to 
name a few) are still unsolved: we must postulate that some unknown 
$^3$He-destruction mechanism is at work in not less than 90\% of low-mass stars 
in order not to overproduce $^3$He in the course of Galactic evolution (see 
discussion in Sect.~\ref{sec:intro}). But which is the physical mechanism 
responsible for that? In a coherent picture, one must also be able to explain 
the existence of a few planetary nebulae (PNe) with high $^3$He content, 
consistent with predictions from standard stellar models (e.g., \cite{b99}). 
Recently, the inclusion of thermohaline mixing in detailed stellar evolutionary 
models (\cite{cz07}) has shown that this is likely to be the `extra mixing' 
mechanism we have been searching for years. Indeed, thermohaline mixing is able 
to efficiently destroy $^3$He on the RGB of low-mass (1--2~M$_\odot$) stars, 
when not inhibited by magnetic fields. Details on this interesting process can 
be found in the contribution by N.~Lagarde and C.~Charbonnel to these 
conference proceedings (see also text by R.~Stancliffe, this volume). The 
expected output of stellar models including thermohaline mixing are complete 
grids of yields. New GCE models computed with such new yields will hopefully 
lead the so-called $^3$He problem to an end.

\section{Helium-4}
\label{sec:4he}

As far as $^4$He is concerned, there are many open issues to be discussed.

\begin{figure}[b]
  \begin{center}
  \includegraphics[height=10cm]{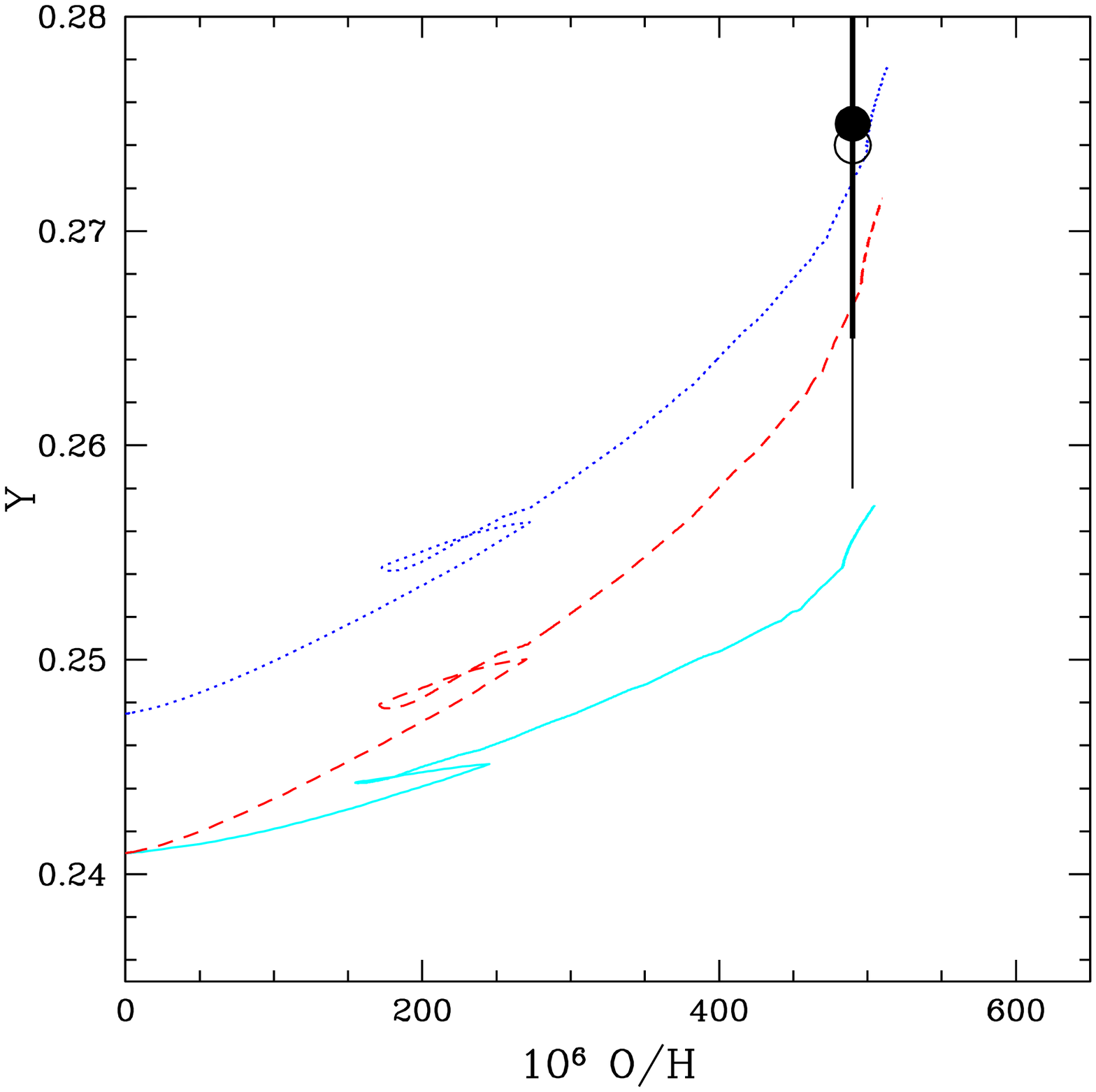} 
  \caption{$Y$ versus 10$^6$ (O/H) in the solar neighbourhood predicted by the 
    \emph{two-infall} model of \cite{c97} with different prescriptions on the 
    stellar nucleosynthesis. Solid line: the model adopts the \cite{vdhg97} 
    yields for low- and intermediate-mass stars and the \cite{ww95} yields for 
    massive stars; dashed line: the model is computed with the yields of 
    \cite{mm02}, taking into account the effects of rotation on stellar 
    evolution, for the whole range of stellar masses; dotted line: same model 
    as the previous one, but starting from a higher primordial $^4$He 
    abundance, $Y_{\mathrm{P}}$~= 0.248 rather than 0.241. The model predictions 
    are compared with the solar value (oxygen from \cite{ap01}, $Y$ from 
    \cite{ag89} -- open circle with thin errorbar -- and \cite{gs98} -- filled 
    circle with thick errorbar). Figure from Chiappini et al. (2003).}
  \label{fig:cris}
 \end{center}
\end{figure}

First of all, it is worth stressing that, when dealing with the evolution of 
$^4$He in the Galactic disc, we can rely on only a few data (see also 
M.~Peimbert et al., this volume), namely, the abundance of $^4$He in the Sun at 
the time of its birth, which is by now quite well established (\cite{a09}), the 
abundance of $^4$He in M\,17, an H\,{\small II} region located in the inner 
Galaxy, and the helium-to-metal enrichment ratio as derived from nearby K-dwarf 
stars. The latter quantity, however, is affected by a rather large error and 
can only be trusted around solar metallicity (see \cite{c07} and contribution 
by L.~Casagrande, this volume). 

Fig.~\ref{fig:cris} shows the behaviour of $Y$ as a function of metallicity 
[10$^6$ (O/H)] in the solar neighbourhood, as predicted by the two-infall model 
of \cite{c97}. The model has been computed with different prescriptions on the 
stellar yields and primordial mass fraction of $^4$He (see figure caption for 
details). Although the models adopting the $^4$He yields from rotating, 
mass-losing stellar models (Fig.~\ref{fig:cris}, dotted and dashed lines) 
provide a good fit to the available solar neighbourhood data, performing 
definitely better than the model using stellar yields without stellar rotation 
and mass loss (Fig.~\ref{fig:cris}, solid line), it has to be reminded that the 
data for M\,17 cannot be satisfactorily reproduced by any GCE model yet (see 
M.~Peimbert et al., this volume, their figure~1).

Another debated topic is the extreme He enhancement in Galactic globular 
clusters (GCs). The presence of multiple MSs in the GCs $\omega$\,Cen and 
NGC\,2808 (\cite{pi05}, 2007), indeed, is most convincingly explained in terms 
of an extreme He enhancement of the bluest MS population (\cite{n04}). Helium 
abundances as high as $Y \sim$~0.4 are suggested, which for $\omega$\,Cen imply 
a helium-to-metal enrichment ratio $\Delta Y/\Delta Z \ge$~70 (\cite{pi05}). 
Such a value is outstandingly larger than that quoted for the solar 
neighbourhood around and above solar metallicity from a sample of nearby 
K-dwarf stars, $\Delta Y/\Delta Z$~= 2.1~$\pm$ 0.9 (\cite{c07}). Attemps have 
been made to explain such extreme $^4$He abundances in the framework of two 
main competing scenarios, the so-called `asymptotic giant branch (AGB) 
self-pollution scenario' (P.~Ventura, this volume) and the so-called `fast 
rotating massive star (FRMS) self-pollution scenario' (T.~Decressin, this 
volume). Both scenarios have to reproduce other chemical peculiarities of GC 
stars besides the `anomalous' $^4$He abundances, and both have advantages and 
disadvantages. As a common limitation, they are presently able to deal only 
with two-population clusters. However, in $\omega$\,Cen -- the object for which 
the most compelling evidence for the need for a \emph{huge} helium enrichment 
is found -- there is clear-cut evidence of the presence of complex, multiple 
populations (e.g., \cite{p00}). We (\cite{r10}) have recently proposed that 
the presence of extreme He-rich stars in $\omega$\,Cen can be explained in the 
context of a model where the cluster is the remnant of a much more massive 
parent system, that evolved in isolation for a relatively long time -- 3~Gyr, 
with the bulk of the stars forming during the first 1~Gyr. The system was then 
captured and partially disrupted by the Milky Way (see contribution by 
D.~Romano et al., this volume). The key ingredient in our model is the 
development of a differential galactic wind, which selectively removes from 
the progenitor galaxy mostly the elements restored to the ISM through fast 
polar winds from massive stars and supernova (SN) explosions. Elements restored 
to the ISM through gentle winds from both AGB and FRMSs are, instead, mostly 
retained in the cluster potential well, where they enter the formation of 
successive stellar generations. Since, according to the latest stellar 
evolutionary computations, $^4$He is dispersed in the ISM by means of 
low-energy stellar winds by both AGBs and FRMSs, while metals are mostly 
expelled through SN explosions, a high $\Delta Y/\Delta Z$ is naturally 
obtained in the framework of our model. Other important observational 
constraints can also be satisfactorily reproduced.

Since differential galactic winds do, as a matter of fact, modify somewhat 
arbitrarily the true (effective) yields of the various elements, a better 
assessment of the stellar yields of $^4$He is mandatory.

Finally, it is worth reminding that the usually quoted value of $Y \sim$~0.4 
for the extreme He-rich GC stars could be revised downwards to as low as $Y 
\sim$~0.3 (L.~Casagrande, this volume).

\section{Conclusions}
\label{sec:conc}

We have reviewed the evolution of the light elements D, $^3$He and $^4$He in 
the Milky Way, emphasizing recent developments and open problems. We summarize 
our conclusions as follows:
\begin{enumerate}
\item GCE models are consistent with the 
  relatively high value of (D/H)$_\mathrm{LISM}$~= (2.0~$\pm$ 0.1)~$\times$ 
  10$^{-5}$ suggested by Prodanovi\'c et al. (2009) from their Bayesian 
  analysis of \emph{FUSE} data. Standard GCE models are instead unable to 
  explain values of (D/H)$_\mathrm{LISM}$ significantly higher/lower than this.
\item The need for some 'extra mixing' to destroy $^3$He in most ($>$ 90\%) of 
  1-–2 M$_\odot$ stars came from GCE arguments more than 30 years ago. The way 
  to the understanding of the physical processes underlying this 
  assumption has been a long one, but now thermohaline mixing seems to be a 
  good candidate to solve the long-standing issue of $^3$He evolution. Stellar 
  yields taking this mechanism into account are needed for use in GCE 
  models.
\item It is still debated whether the chemical peculiarities seen in a fraction 
  of Galactic globular cluster stars -- first of all an impressive helium 
  enrichment -- are due to self-pollution from AGBs or FRMSs. In the very 
  peculiar case of $\omega$\,Cen, which likely suffered a complicated star 
  formation history as the nucleus of a larger system, both stellar categories 
  should have polluted the ISM. In such a scenario, the observed chemical 
  `anomalies' would be driven by the action of differential galactic outflows, 
  venting out preferentially metals. However, it is still unclear if such an 
  extreme scenario could apply to other GCs as well.
\end{enumerate}

\section*{Acknowledgements}

I thank the organizers for their kind invitation and for giving me the 
opportunity to attend such a lively conference. I would like to express my 
gratitude to Francesca Matteucci and Monica Tosi for advice over the years, and 
to Johannes Geiss and Gary Steigman for helpful discussions. Generous financial 
support from IAU is also gratefully acknowledged. The author's research at 
Bologna University is supported by Italian MIUR under grant PRIN\,2007, 
prot.~2007JJC53X\_001.

\end{document}